\documentclass[letterpaper, 10 pt, conference]{ieeeconf}  

\IEEEoverridecommandlockouts

\title{\LARGE \bf
Deep Neural Networks for Multiple Speaker Detection and Localization
}

\author{Weipeng He$^{1,2}$, Petr Motlicek$^{1}$ and Jean-Marc Odobez$^{1,2}$%
\thanks{*This research has been partially funded by the European Union’s Horizon 2020 research and innovation programme under grant agreement no. 688147 (MuMMER, mummer-project.eu)}%
\thanks{$^{1}$Idiap Research Institute, Switzerland. \texttt{\{weipeng.he, petr.motlicek, odobez\}@idiap.ch}}%
\thanks{$^{2}$\'Ecole Polytechnique F\'ed\'erale de Lausanne (EPFL), Switzerland.}%
}

\usepackage{mypre}

\begin{document}

\maketitle
\thispagestyle{empty}
\pagestyle{empty}

\begin{abstract}
We propose to use neural networks for simultaneous detection and localization of multiple sound sources in human-robot interaction. In contrast to conventional signal processing techniques, neural network-based sound source localization methods require fewer strong assumptions about the environment. Previous neural network-based methods have been focusing on localizing a single sound source, which do not extend to multiple sources in terms of detection and localization. In this paper, we thus propose a likelihood-based encoding of the network output, which naturally allows the detection of an arbitrary number of sources. In addition, we investigate the use of sub-band cross-correlation information as features for better localization in sound mixtures, as well as three different network architectures based on different motivations. Experiments on real data recorded from a robot show that our proposed methods significantly outperform the popular spatial spectrum-based approaches.

\end{abstract}


\section{Introduction}

\subsection{Motivation}

\begin{figure}[t]
  \centering
  \includegraphics[height=.3\linewidth]{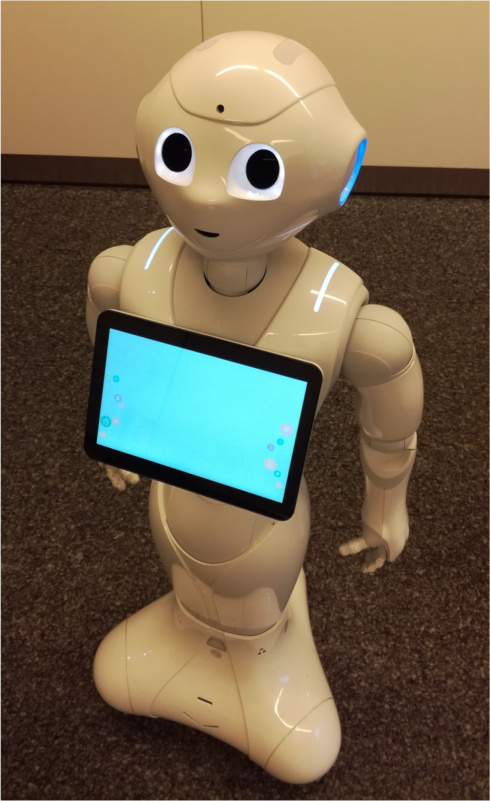}\hspace{1.2cm}\includegraphics[height=.3\linewidth]{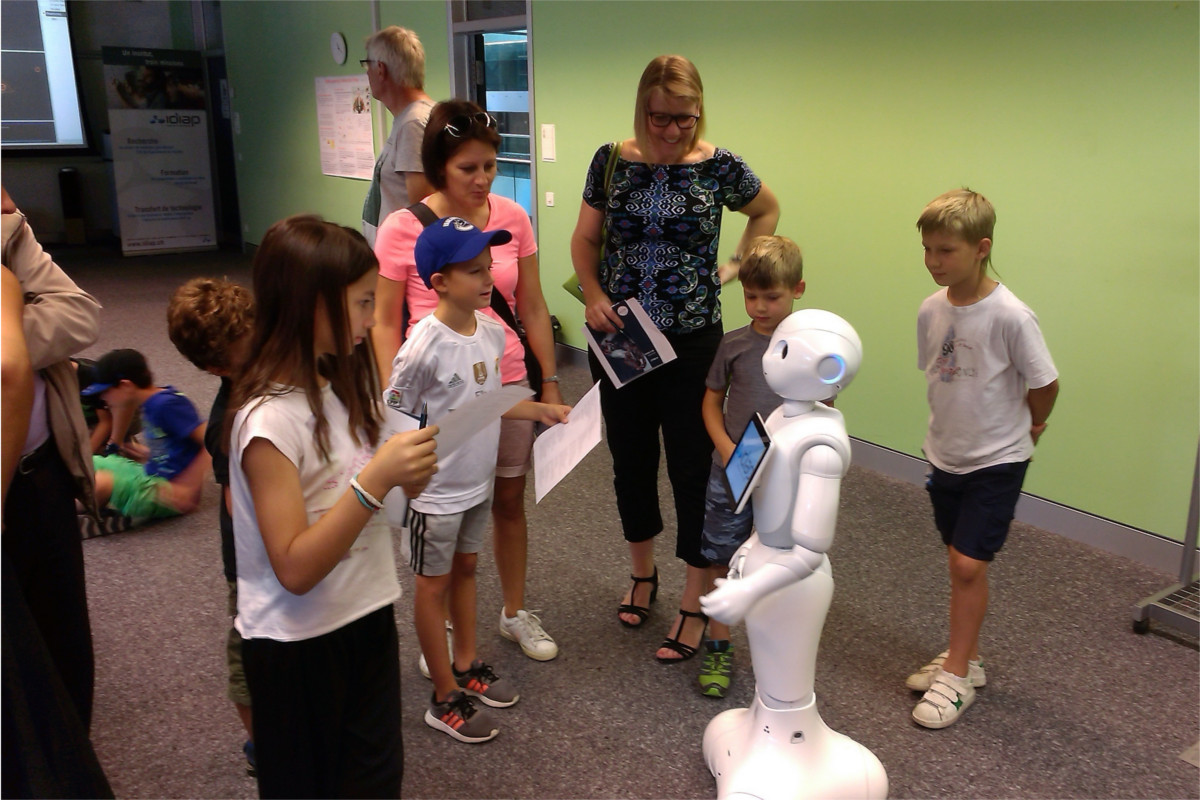}
  \caption{The robot, Pepper, used in our experiments and a typical HRI scenario where the robot interacts with multiple persons.}\label{fig:hri_scene}
\end{figure}

Sound source localization (SSL) and speaker detection are crucial components in multi-party human-robot interaction (HRI), where the robot needs to precisely detect where and who the speaker is and responds appropriately (Fig.~\ref{fig:hri_scene}). In addition, robust output from SSL is essential for further HRI analysis (e.g.\ speech recognition, speaker identification, etc.) which provides a reliable source of information to be combined with other modalities towards improved HRI\@. Although SSL has been studied for decades, it is still a challenging topic in real HRI applications, due to the following conditions:
\begin{itemize}
  \item Noisy environments and strong robot ego-noise;
  \item Multiple simultaneous speakers;
  \item Short and low-energy utterances, as responses to questions or non-verbal feedback;
  \item Obstacles such as robot body blocking sound direct path.
\end{itemize}

Traditionally, SSL is considered a signal processing problem. The solutions are analytically derived with assumptions about the signal, noise and environment~\cite{knapp_generalized_1976,schmidt_multiple_1986,brandstein_robust_1997}. However, many of the assumptions do not hold well under the above-mentioned conditions, which may severely impact their performance. Alternatively, researchers have recently adopted machine learning approaches with neural networks (NN). Indeed, with a sufficient amount of data, the NNs can in principle learn the mapping from the localization cues to the direction-of-arrival (DOA) without making strong assumptions. Surprisingly, most of the learning-based methods do not address the problem of multiple sound sources and in particular, the simultaneous detection and localization of multiple voices in real multi-party HRI scenarios have not been well studied.

\subsection{Existing Neural Network-based SSL Methods}
Although the earliest attempts of using neural networks for SSL date back to the 1990s~\cite{yuhas_automated_1992,datum_artificial_1996}, it was not until recently that researchers started to pay more attention to such learning-based approaches. With the large increase of computational power and advances in deep neural networks (DNN), several methods were shown to achieve promising single SSL performance~\cite{youssef_learning-based_2013,ma_exploiting_2015,xiao_learning-based_2015,takeda_sound_2016,yalta_sound_2017}. Nevertheless, most of these methods aim at detecting only one source, focusing the research on the localization accuracy. In particular, they formulate the problem as the classification of an audio input into one ``class'' label associated with a location, and optimizing the posterior probability of such labels. Unfortunately, such posterior probability encoding cannot be easily extended to multiple sound source situations.

Neural networks that are trained for localizing a single source can be applied for multi-source localization by pooling the network outputs (i.e.\ posterior probabilities) over multiple time frames~\cite{ma_exploiting_2015}. However, this method requires a known number of sources and a long period of time as input for pooling. Such limitations make it not practical for real applications.

Localization of two sources is addressed in~\cite{takeda_discriminative_2016}, which encodes the output as two marginal posterior probability vectors. However, an ad-hoc location-based ordering is introduced to decide the source-to-vector assignment, rendering the posteriors dependent on each other and the encoding somewhat ambiguous. That is, the same source may need to be predicted as the first source if alone, or as the second one if another signal with a preceding label is present.

\begin{table*}[t]
  \caption{Comparison of our methods with existing NN-based SSL approaches}\label{tab:nn_ssl}
  \centering
  \begin{tabular}{r@{\hspace{.8cm}}c@{\hspace{.8cm}}c@{\hspace{.8cm}}c@{\hspace{.8cm}}c}
    \toprule
    Approach & Number of Sources & Input Size & Input Feature & Output Coding \\
    \midrule
    Datum et al.~\cite{datum_artificial_1996}       & 1            & -         & IPD and ITD per freq. & Gaussian-shaped function \\
    Xiao et al.~\cite{xiao_learning-based_2015}     & 1            & Utterance & GCC-PHAT coefficients & Posterior probability \\
    Takeda et al.~\cite{takeda_sound_2016}          & 0 or 1       & 200ms     & MUSIC eigenvectors & Posterior probability \\
    Yalta et al.~\cite{yalta_sound_2017}            & 0 or 1       & 200ms     & Power spectrogram & Posterior probability \\
    Ma et al.~\cite{ma_exploiting_2015}             & Known multiple & Utterance & CCF and ILD per freq. & Posterior probability \\
    Takeda et al.~\cite{takeda_discriminative_2016} & 0, 1, 2      & 200ms     & MUSIC eigenvectors & Marginal posterior probability \\
    \cmidrule(r){1-5}
    Ours & Unknown multiple & 170ms & GCC-PHAT and GCCFB & Likelihood-based coding \\
    \bottomrule
  \end{tabular}
\end{table*}

\subsection{Contributions}

This paper investigates NN-based SSL methods applied to real HRI scenarios (Fig.~\ref{fig:hri_scene}). In contrast to previous studies (Table~\ref{tab:nn_ssl}), the methods are required to cope with short input, overlapping speech, an unknown number of sources and strong ego-noise. We emphasize their application in real conditions by testing the methods with recorded data from the robot Pepper\footnote{\url{http://doc.aldebaran.com/2-5/home_pepper.html}}.

In this paper, we propose three NN architectures for multiple SSL based on different motivations. The NNs adopt a likelihood-based output encoding that can handle an arbitrary number of sources. And, we investigate the employment of sub-band cross correlation information as an input feature for better localization cues in speech mixtures. The experiments indicate that the proposed methods significantly outperform the baseline methods.

Furthermore, we collect and release a benchmark dataset\footnote{\url{https://www.idiap.ch/dataset/sslr/}} of real recordings for developing and evaluating learning-based SSL in HRI\@.

\section{Proposed Method}
In this section, we describe our proposed NN models for multiple SSL\@. We consider the localization of sounds in the azimuth direction in individual time frames, which are 170ms long. We denote the number of sources by $N$ and the number of microphones by $M$. The input signal is represented by the short time Fourier transforms (STFT): $X_i(\omega), i=1,\dots,M$, where $i$ is the microphone index and $\omega$ is the frequency in the discrete domain. We omit the time index for clarity, because none of the methods described below exploit context information or temporal relations.

\subsection{Input Features}

The generalized cross-correlation with phase transform (GCC-PHAT)~\cite{knapp_generalized_1976} is the most popular method for estimating the time difference of arrival (TDOA) between microphones, which is an important clue for SSL\@. Here, we use two types of features based on GCC-PHAT\@ at frame level.

\textbf{GCC-PHAT coefficients:}
The first type of input feature is represented by the center GCC-PHAT values of all $M(M-1)/2$ microphone pairs as used in~\cite{xiao_learning-based_2015}. The GCC-PHAT between channel $i$ and $j$ is formulated as:
\begin{equation}
  g_{ij}(\tau) = \sum_{\omega} \mathcal{R}\left(\frac{X_i(\omega) X_j{(\omega)}^*}{\left|X_i(\omega) X_j{(\omega)}^*\right|} e^{j \omega \tau}\right),
\end{equation}
where $\tau$ is the delay in the discrete domain, ${(\cdot)}^*$ denotes the complex conjugation, and $\mathcal{R}(\cdot)$ denotes the real part of a complex number. The peak in GCC-PHAT is used to estimate the TDOA\@. However, under real condition, the GCC-PHAT is corrupted by noise and reverberation. Therefore, we use the full GCC-PHAT function as the input feature instead of a single estimation of the TDOA\@. In our experiments, we use the center 51 delays ($\tau \in [-25,25]$).

\begin{figure}[t]
  \centering \sf \tiny \mathversion{sfnums}
  \includegraphics[width=\linewidth]{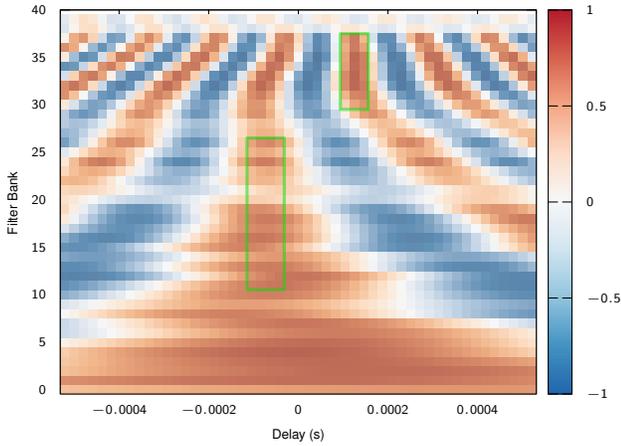}
  \caption{Example of GCCFB extracted from a frame with two overlapping sound sources.}\label{fig:gccfb}
\end{figure}

\textbf{GCC-PHAT on mel-scale filter bank:}
The GCC-PHAT is not optimal for TDOA estimation of multiple source signals since it equally sums over all frequency bins disregarding the ``sparsity'' of speech signals in the time-frequency (TF) domain and the randomly distributed noise which may be stronger than the signal in some TF bins. To preserve delay information on each frequency band and to allow sub-band analysis, we propose to use GCC-PHAT on mel-scale filter bank (GCCFB). Hence, the second type of input feature is formulated as:
\begin{equation}
  g_{ij}(f, \tau) = \frac{\sum_{\omega \in \Omega_f} \mathcal{R}\left(H_f(\omega) \frac{X_i(\omega) {X_j(\omega)}^*}{\left|X_i(\omega) X_j{(\omega)}^*\right|} e^{j \omega \tau}\right)}{\sum_{\omega \in \Omega_f} H_f(\omega)},
\end{equation}
where $f$ is the filter index, $H_f$ is the transfer function of the $f$-th mel-scaled triangular filter, and $\Omega_f$ is the support of $H_f$. Fig.~\ref{fig:gccfb} shows an example of the GCCFB of a frame where two speech signals overlap. Each row corresponds to the GCC-PHAT in an individual frequency band. The frequency-based decomposition allows the estimation of the TDOAs by looking into local areas rather than across all frequency bins. In the example, the areas marked by the green rectangles correspond to two separate sources with different delays and which produce high cross-correlation values in different frequency bands (and hence, for different filter indices). In the experiments, we use 40 mel-scale filters covering the frequencies from 100 to 8000 Hz.

\subsection{Likelihood-based Output Coding}
\begin{figure}[t]
  \centering \tiny \sffamily \mathversion{sfnums}
  \includegraphics[width=\linewidth]{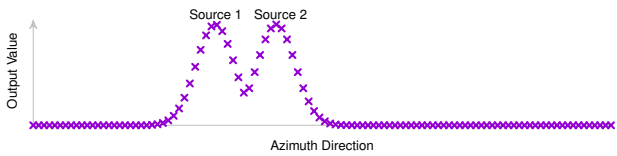}
  \caption{Output coding for multiple sources.}\label{fig:outcode}
\end{figure}

\textbf{Encoding:}
We design the multiple SSL output coding as the likelihood of a sound source being in each direction. Specifically, the output is encoded into a vector $\{o_i\}$ of 360 values, each of which is associated with an individual azimuth direction $\theta_i$. The values are defined as the maximum of Gaussian-like functions centered around the true DOAs:
\begin{equation}
  o_i = \begin{cases}
    \max_{j=1}^{N} \left\{ e^{-d{(\theta_i,\theta^{(s)}_j)}^2 / \sigma^2} \right\} & \quad \text{if } N > 0 \\
                                                                               0 & \quad \text{otherwise} \\
        \end{cases},
\end{equation}
where $\theta^{(s)}_j$ is the ground truth DOA of the $j$-th source, $\sigma$ is the value to control the width of the Gaussian-like curves and $d(\cdot,\cdot)$ denotes the angular distance. The output coding resembles a spatial spectrum, which is a function that peaks at the true DOAs (Fig.~\ref{fig:outcode}).

Unlike posterior probability coding, the likelihood-based coding is not constrained as a probability distribution (the output layer is not normalized by a softmax function). It can be all zero when there is no sound source, or contains $N$ peaks when there are $N$ sources. The coding can handle the detection of an arbitrary number of sources. In addition, the soft assignment of the output values, in contrast to the $0/1$ assignment in posterior coding, takes the correlation between adjacent directions into account allowing better generalization of the neural networks.

\textbf{Decoding:}
During the test phase, we decode the output by finding the peaks that are above a given threshold $\xi$:
\begin{equation}\label{eq:decode}
  \text{Prediction} = \left\{ \theta_i : o_i > \xi \quad \text{and} \quad o_i = \max_{d(\theta_j, \theta_i) < \sigma_n} o_j \right\},
\end{equation}
with $\sigma_n$ being the neighborhood distance. We choose $\sigma = \sigma_n = 8\text{\textdegree}$ for the experiments.

\subsection{Neural Network Architectures}
We investigate three different types of NN architectures for sound source localization.

\begin{figure}[t]
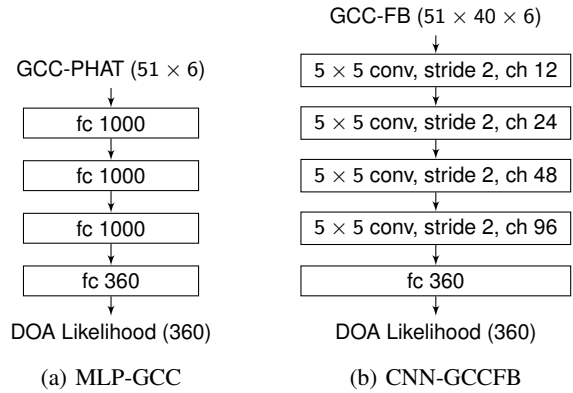

  \begin{subfigure}[b]{0.4\linewidth}
    \centering \sf \footnotesize \mathversion{sfnums}
    \includegraphics{arch_mlp.tikz}
    \caption{MLP-GCC}\label{fig:arch_mlp}
  \end{subfigure}%
  \begin{subfigure}[b]{0.6\linewidth}
    \centering \sf \footnotesize \mathversion{sfnums}
    \includegraphics{arch_cnn.tikz}
    \caption{CNN-GCCFB}\label{fig:arch_cnn}
  \end{subfigure}%
  \caption{Two neural network architectures for multiple SSL.}
\end{figure}

\textbf{MLP-GCC} (Multilayer perceptron with GCC-PHAT): As illustrated in Fig.~\ref{fig:arch_mlp}, the MLP-GCC uses GCC-PHAT as input and contains three hidden layers, each of which is a fully-connected layer with a rectified linear unit (ReLU) activation function~\cite{nair_rectified_2010} and batch normalization (BN)~\cite{ioffe_batch_2015}. The last layer is a fully connected layer with sigmoid activation function. The sigmoid function is bounded between $0$ and $1$, which is the range of the desired output. According to our experiments, this helps the network to converge to a better result.

\textbf{CNN-GCCFB} (Convolutional neural network with GCCFB):
Fully connected NNs are not suitable for high-dimensional input features (such as GCCFB) because the large dimension introduces a large amount of parameters to be trained, making the network computationally expensive and prone to overfitting. Convolutional neural networks (CNN) can learn local features with reduced amount of parameters by using weight sharing. This leads to the idea of using CNN for the input feature of GCCFB\@.

We use the CNN structure shown in Fig.~\ref{fig:arch_cnn}, which consists of four convolutional layers (with ReLU activation and BN) and a fully connected layer at the output (with sigmoid activation). The local features are not shift invariant since the position of the feature (the delay and frequency) is the important cue for SSL\@. Therefore, we do not apply any pooling after convolution. Instead, we apply the filters with a stride of 2, expecting that the network learns its own spatial downsampling.

\textbf{TSNN-GCCFB} (Two-stage neural network with GCCFB):
The CNN-GCCFB considers the input features as images without taking their properties into account, which may not yield the best model. Thus, for the third architecture, we design the weight sharing in the network with the knowledge about the GCCFB\@:
\begin{itemize}
  \item In each TF bin, there is generally only one predominant speech source, thus we can do analysis or implicit DOA estimation in each frequency band before such information is aggregated into a broadband prediction.
  \item Features with the same delay on different microphone pairs do not correspond to each other locally. Instead, feature extraction or filters should take the whole delay axis into account.
\end{itemize}

\begin{figure}[t]
  \centering \sf \footnotesize \mathversion{sfnums}
  \includegraphics{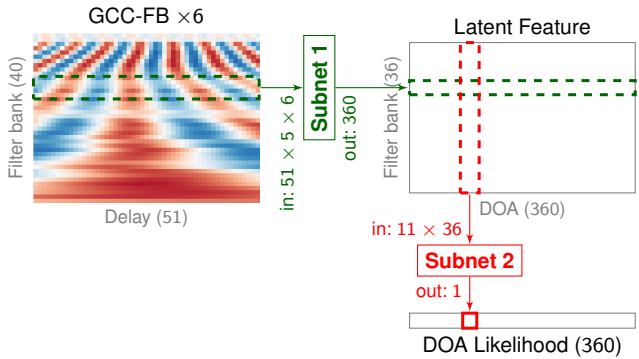}
  \caption{NN architecture of two-stage neural network with GCCFB as input. The first and second stages are marked as green and red, respectively.}\label{fig:arch_tsrnn}
\end{figure}

Based on these considerations, we propose the two-stage neural network (Fig.~\ref{fig:arch_tsrnn}). The first stage extracts latent DOA features in each filter bank, by repeatedly applying \emph{Subnet 1} on individual frequency regions that span all delays and all microphone pairs. The second stage aggregates information across all frequencies in a neighbor DOA area and outputs the likelihood of a sound being in each DOA\@. Similarly, the \emph{Subnet 2} is repeatedly used for all DOAs in the second stage. To train such network, we adopt a two-step training scheme: First, we train the Subnet 1 in the first stage using the DOA likelihood as the desired latent feature. In such way, we obtain DOA and frequency-related features that help the NN to converge to a better result in the next step. During the second step, both stages are trained in an end-to-end manner. In our experiments, Subnet 1 is a 2-hidden-layer MLP, and Subnet 2 is a 1-hidden-layer MLP\@. All the hidden layers are of size 500.

\section{Experiment}
We implemented the proposed methods and compared them to the traditional SSL approaches with the data collected from a robot.

\subsection{Datasets}
For the development and evaluation of learning-based SSL methods, we collected two sets of real data: one with loudspeaker and the other with human subjects (see Table~\ref{tab:datasets}).

We use Pepper for the recording of both sets. There are four microphones on the top of its head, forming a rectangle of $5.8 \times 6.9$ cm. The microphones are directional with a forward look direction. The audio signals received by the microphones are strongly affected by the robot's fan noise from inside the head. The sample rate is 48 kHz.

\begin{figure}[t]
  \centering
  \begin{subfigure}[b]{0.45\linewidth}
    \centering
    \includegraphics[width=\linewidth]{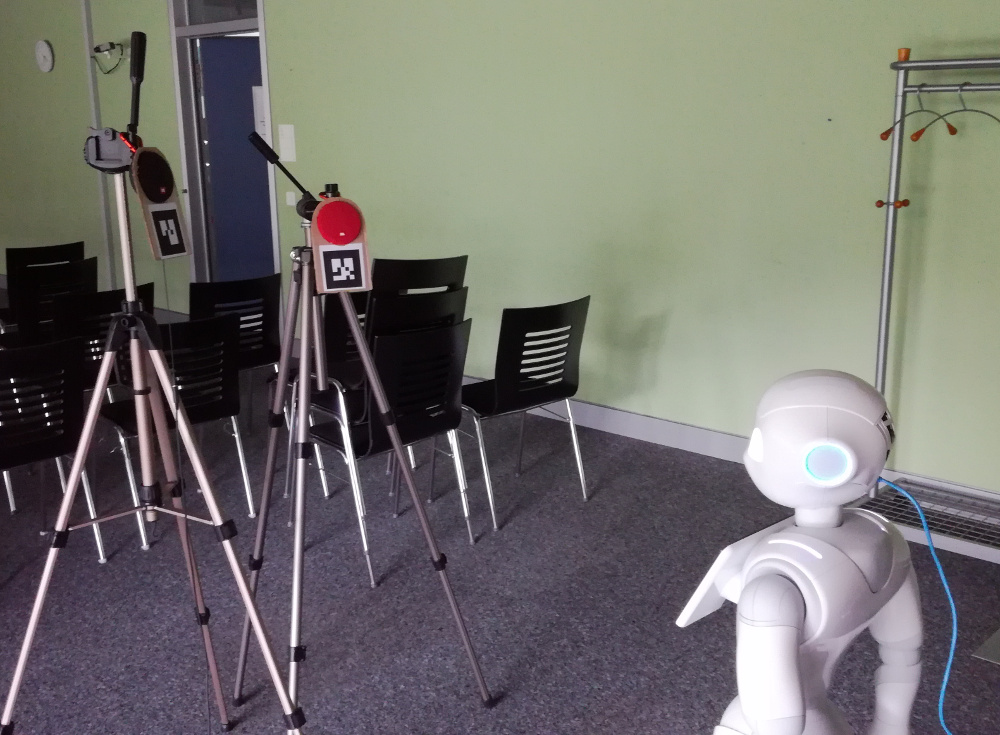}
    \caption{Loudspeakers.}\label{fig:collection_room}
  \end{subfigure}
  \hspace{1em}
  \begin{subfigure}[b]{0.45\linewidth}
    \includegraphics[width=\linewidth]{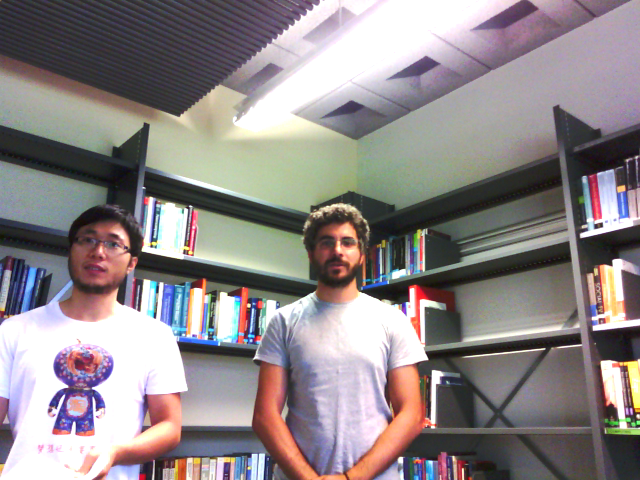}
    \caption{Human subjects.}\label{fig:human_subjects}
  \end{subfigure}
  \caption{Data collection with Pepper.}\label{fig:ssl_recording}
\end{figure}

\textbf{Recording with loudspeakers:}
We collected data by recording clean speech played from loudspeakers (Fig.~\ref{fig:collection_room}). The clean speech data were selected from the AMI corpus~\cite{mccowan_ami_2005}, which contains spontaneous speech of people interacting in meetings. The loudspeakers were attached with markers so that they can be automatically located by the camera on the robot. The data were recorded in rooms of different sizes, with the robot and loudspeakers put at random places. We programmed the robot to move its head automatically to acquire a large diversity of loudspeaker-to-robot positions.

\textbf{Recording with human subjects:}
To evaluate SSL methods in real HRI, we collected the second dataset that involves human subjects (Fig.~\ref{fig:human_subjects}). During the recording, the subjects spoke to the robot with phrases for interactions. This dataset includes recordings with single utterances as well as overlapping ones. We manually annotated the voice activity detection (VAD) labels and automatically acquired the mouth position by running a multiple person tracker~\cite{khalidov_real-time_2017} with detection from the convolutional pose machine (CPM)~\cite{wei_convolutional_2016}.

\begin{table}[t]
  \caption{Specifications of the recorded data}\label{tab:datasets}
  \centering
    \begin{tabular}{lccc}
      \toprule
                            & \multicolumn{2}{c}{Loudspeaker} & Human         \\
                            \cmidrule(r){2-3} \cmidrule(r){4-4}
                            & Training      & Test            & Test          \\
      \midrule
      \# of files           & $4208$        & $2393$          & $21$          \\
\hspace{1em}- single source & $2808$        & $1597$          & $-$           \\
\hspace{1em}- two sources   & $1400$        & $796$           & $21$          \\
      \# of male speakers   & $105$         & $8$             & $12$          \\
      \# of female speakers & $43$          & $8$             & $2$           \\
      Total duration        & 16 hours      & 8 hours         & 220 seconds   \\
      Azimuth (\textdegree) & $[-180, 180]$ & $[-180, 180]$   & $[-24, 23]$   \\
    Elevation (\textdegree) & $[-39, 56]$   & $[-29, 45]$     & $[-14, 13]$   \\
      Distance (m)          & $[0.5, 1.8]$  & $[0.5, 1.9]$    & $[0.8, 2.1]$  \\
      \bottomrule
    \end{tabular}
\end{table}

\subsection{Evaluation Protocol}

We evaluate multiple SSL methods at frame level under two different conditions: the number of sources is known or unknown.

\textbf{Known number of sources:} We select the $N$ highest peaks of the output as the predicted DOAs and match them with ground truth DOAs one by one, and we compute the \emph{mean absolute error} (MAE). In addition, we consider the \emph{accuracy} (ACC) as the percentage of correct predictions. By saying a prediction is correct, we mean the error of the prediction is less than a given admissible error $E_a$.

\textbf{Unknown number of sources:} We consider the ability of both detection and localization. To do this, we make predictions based on Eq.~\ref{eq:decode}, and compute the \emph{precision} vs.\ \emph{recall} curve by varying the prediction threshold $\xi$. The precision is the percentage of correct predictions among all predictions. And, the recall is the percentage of correct detection out of all ground truth sources.

\subsection{Network Training}
We trained the NN with the loudspeaker training set, which includes a total of 506k frames of no source, one source, or two sources. We used the Adam optimizer~\cite{kingma_adam:_2014} with mean squared error (MSE) loss and mini-batch size of 256. The MLP-GCC and CNN-GCCFB were trained for ten epochs. We trained the TSNN-GCCFB for four epochs for the first stage and another ten epochs for the end-to-end training.

\subsection{Baseline Methods}
We include the following popular spatial spectrum-based methods for comparison:
\begin{itemize}
  \item SRP-PHAT\@: steered response power with phase transform~\cite{brandstein_robust_1997};
  \item SRP-NONLIN\@: SRP-PHAT with a non-linear modification of the score, it is a multi-channel extension of GCC-NONLIN from~\cite{blandin_multi-source_2012};
  \item MVDR-SNR\@: minimum variance distortionless response (MVDR) beamforming~\cite{krim_two_1996} with signal-to-noise ratio (SNR) as score~\cite{blandin_multi-source_2012};
  \item SEVD-MUSIC\@: multiple signal classification (MUSIC)~\cite{schmidt_multiple_1986}, assuming spatially white noise and one signal in each bin;
  \item GEVD-MUSIC\@: MUSIC with generalized eigenvector decomposition~\cite{schmidt_multiple_1986,nakamura_intelligent_2009}, assuming noise is pre-measured and one signal in each TF bin.
\end{itemize}

For all the above methods, the empirical spatial covariance matrices are computed using blocks containing 7 frames of 2048 samples with 50\% overlap, so that each block is 170ms long.

\subsection{Results}

Table~\ref{tab:perf_known} shows the results of localization with a known number of sources. On the loudspeaker dataset, all three proposed NN models achieve on average less than 5\textdegree\ error and more than 90\% accuracy, while the best baseline method, (SRP-PHAT) has 21.5\textdegree\ error and only 78\% accuracy. For the human subject dataset, the baseline methods have slightly better MAE on frames with a single source. However, the proposed methods outperform the baseline methods in terms of accuracy, especially on frames with overlapping sources. Note that, the loudspeaker dataset is in general more challenging because it contains samples with lower SNR and wider range of azimuth directions. The sources from the rear are difficult to detect due to the directivity of the microphones.

\begin{table*}[t]
  \caption{Performance assuming a known number of sources. $E_a=5\text{\textdegree}$.}\label{tab:perf_known}
  \centering
    \begin{tabular}{lcccccccccccc}
      \toprule
      Dataset & \multicolumn{6}{c}{Loudspeaker} & \multicolumn{6}{c}{Human} \\
      \cmidrule(r){2-7} \cmidrule(r){8-13}
      Subset (\# of frames) & \multicolumn{2}{c}{Overall (207k)} & \multicolumn{2}{c}{$N=1$ (178k)} & \multicolumn{2}{c}{$N=2$ (29k)}
              & \multicolumn{2}{c}{Overall (929)} & \multicolumn{2}{c}{$N=1$ (788)} & \multicolumn{2}{c}{$N=2$ (141)} \\
      \cmidrule(r){2-3} \cmidrule(r){4-5} \cmidrule(r){6-7} \cmidrule(r){8-9} \cmidrule(r){10-11} \cmidrule(r){12-13}
              & MAE (\textdegree) & ACC & MAE (\textdegree) & ACC & MAE (\textdegree) & ACC & MAE (\textdegree) & ACC & MAE (\textdegree) & ACC & MAE (\textdegree) & ACC \\

      \midrule
      MLP-GCC & $4.89$ & $\bm{0.92}$ & $4.18$ & $\bm{0.94}$ & $9.21$ & $\bm{0.77}$ & $4.99$ & $0.93$ & $4.44$ & $0.94$ & $8.06$ & $0.84$ \\
      CNN-GCCFB   & $\bm{4.80}$ & $0.90$ & $\bm{4.11}$ & $0.93$ & $\bm{9.06}$ & $0.73$ & $4.82$ & $0.93$ & $4.19$ & $\bm{0.96}$ & $8.34$ & $0.77$ \\
      TSNN-GCCFB & $5.41$ & $0.91$ & $4.64$ & $0.93$ & $10.10$ & $\bm{0.77}$ & $\bm{4.14}$ & $\bm{0.95}$ & $3.84$ & $\bm{0.96}$ & $\bm{5.84}$ & $\bm{0.90}$ \\
      \cmidrule{1-13}
      SRP-PHAT~\cite{brandstein_robust_1997}       & $21.51$ & $0.78$ & $19.00$ & $0.82$ & $36.95$ & $0.50$ & $5.39$ & $0.88$ & $2.62$ & $0.93$ & $20.90$ & $0.56$ \\
      SRP-NONLIN~\cite{blandin_multi-source_2012}  & $25.71$ & $0.73$ & $23.77$ & $0.77$ & $37.61$ & $0.51$ & $4.84$ & $0.90$ & $2.47$ & $0.94$ & $18.11$ & $0.68$ \\
      MVDR-SNR~\cite{blandin_multi-source_2012}    & $23.17$ & $0.76$ & $21.22$ & $0.79$ & $35.19$ & $0.55$ & $4.39$ & $0.90$ & $\bm{2.45}$ & $0.94$ & $15.21$ & $0.68$ \\
      SEVD-MUSIC~\cite{schmidt_multiple_1986}      & $29.07$ & $0.66$ & $27.59$ & $0.69$ & $38.14$ & $0.47$ & $6.36$ & $0.85$ & $3.00$ & $0.88$ & $25.14$ & $0.64$ \\
      GEVD-MUSIC~\cite{nakamura_intelligent_2009}  & $25.43$ & $0.64$ & $23.18$ & $0.67$ & $39.28$ & $0.44$ & $6.45$ & $0.81$ & $3.62$ & $0.85$ & $22.24$ & $0.63$ \\
      \bottomrule
    \end{tabular}
\end{table*}

\begin{figure*}[t]
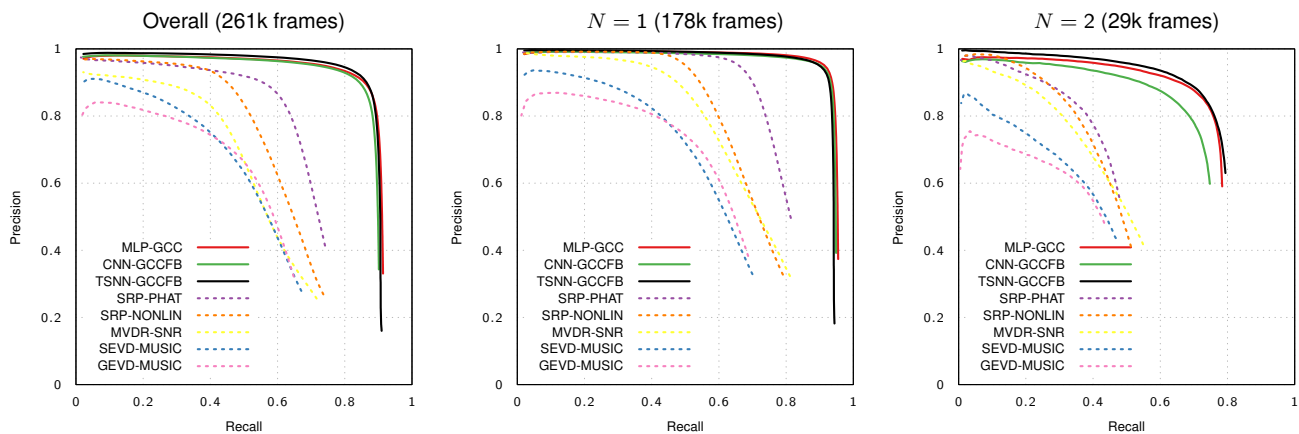

  \centering
  {\tiny \sffamily \mathversion{sfnums}
  \includegraphics[width=.33\linewidth]{pr_lspeaker.tikz}%
  \includegraphics[width=.33\linewidth]{pr_lspeaker_n1.tikz}%
  \includegraphics[width=.33\linewidth]{pr_lspeaker_n2.tikz}
  }
  \caption{Detection and localization performance on recordings with loudspeakers. $E_a=5\text{\textdegree}$.}\label{fig:pr_lspeaker}
\end{figure*}

\begin{figure*}[t]
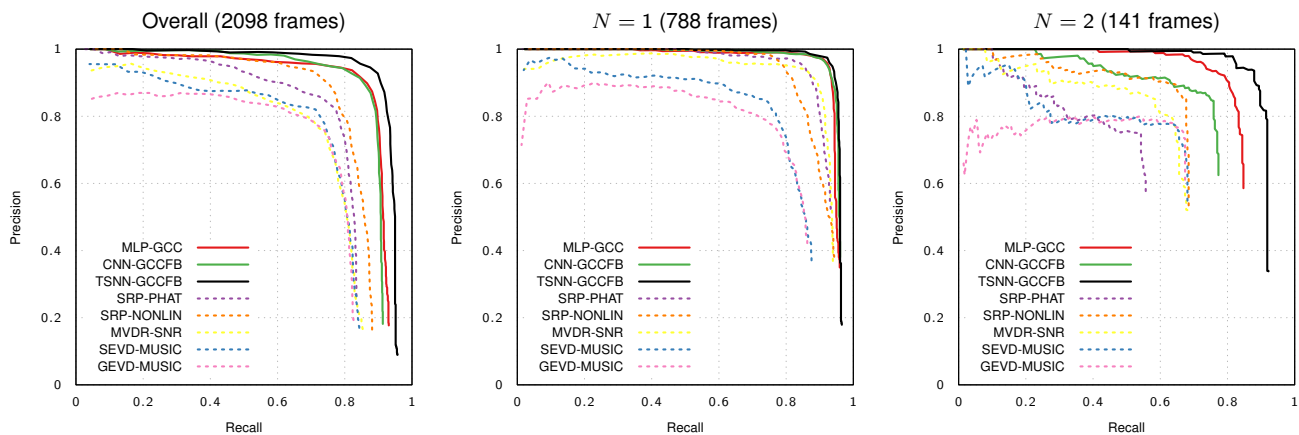

  \centering
  {\tiny \sffamily \mathversion{sfnums}
  \includegraphics[width=.33\linewidth]{pr_human.tikz}%
  \includegraphics[width=.33\linewidth]{pr_human_n1.tikz}%
  \includegraphics[width=.33\linewidth]{pr_human_n2.tikz}
  }
  \caption{Detection and localization performance on recordings with human subjects. $E_a=5\text{\textdegree}$.}\label{fig:pr_human}
\end{figure*}

In terms of simultaneous detection and localization with an unknown number of sources, our proposed methods outperform the baseline methods, achieving approximately 90\% precision and recall on both datasets (Fig.~\ref{fig:pr_lspeaker} and~\ref{fig:pr_human}). Among the three proposed models, the TSNN-GCCFB achieves the best results with its better performance on overlapping frames. This justifies that the usage of the sub-band feature and two-stage structure is beneficial for multiple SSL\@. We also notice that, unlike signal processing approaches, our NN-based methods are not affected by the condition of an unknown number of sources. This indicates that our output coding and data-driven approach are effective for detecting the number of sources. A demonstration video is accompanied with this paper.

\section{Conclusion}
This paper has investigated neural network models for simultaneous detection and localization of speakers. We have proposed a likelihood-based output coding, making it possible to train the NN to detect an arbitrary number of overlapping sound sources. We have collected a large amount of real data, including recordings with loudspeakers and humans, for training and evaluation. The results of the comprehensive evaluation show that our proposed methods significantly outperform the traditional spatial spectrum-based methods.

The current study is potentially limited by the training data samples, which are not likely to cover all possible combinations of source positions, since the number of combinations grows exponentially with the number of sources. Future work will explore network models that can generalize for multiple sound sources with limited training data. We will also explore the robustness of the NN to other more challenging noise, such as cocktail party noise. Furthermore, we will investigate the incorporation of temporal context, which was omitted in our experiments.

\bibliographystyle{IEEEtranNoURL}
\bibliography{audio,vision,dl,mummer}

\end{document}